\documentclass[linenumbers,twocolumn]{aastex631}
\usepackage{listings}
\usepackage{xcolor}
\usepackage[utf8]{inputenc}  
\usepackage{csvsimple}
\usepackage{booktabs}
\usepackage{pgfplotstable}
\usepackage{datatool}
\usepackage[figuresright]{rotating}
\usepackage{makecell}
\usepackage{tabularx,ragged2e}
\usepackage{amsmath}

\definecolor{Dkgreen}{rgb}{0,0.6,0}
\definecolor{Gray}{rgb}{0.5,0.5,0.5}
\definecolor{Mauve}{rgb}{0.58,0,0.82}
\definecolor{Red}{rgb}{1,0,0}
\definecolor{Violet}{rgb}{0.93,0.51,0.93}
\definecolor{Blue}{rgb}{0,0,1}

\begin{document}
\nolinenumbers
\title{Static Microlensing: Concept, Method and Candidates}

\correspondingauthor{Wentao Luo}
    \email{wtluo@ustc.edu.cn,guoqi1@ustc.edu,weily23@mails.tsinghua.edu.cn}
    
\author{Qi Guo}
\affiliation{Department of Astronomy, University of Science and Technology of China, Hefei, Anhui 230026, China}
%\email{guoqi1@ustc.edu}

\author[0009-0003-4545-0945]{Leyao Wei}
\affiliation{School of Physics and Astronomy, Sun Yat-sen University, Zhuhai Campus, \\
2 Daxue Road, Xiangzhou District, Zhuhai, P. R. China}
\affiliation{Department of Astronomy, Tsinghua University, Beijing 100084, China}
 %\email{weily23@mails.tsinghua.edu.cn}

\author[0000-0003-1297-6142]{Wentao Luo}
\affiliation{Department of Astronomy, University of Science and Technology of China, Hefei, Anhui 230026, China}
\affiliation{Institute of Deep Space Sciences, Deep Space Exploration Lab, Hefei, Anhui 230026, China}
 %\email{wtluo@ustc.edu.cn}
 
\author[0009-0000-5381-7039]{Shurui Lin}
\affiliation{Department of Astronomy, University of Science and Technology of China, Hefei, Anhui 230026, China}

\author[0000-0003-3616-6486]{Qinxun Li}
\affiliation{Department of Physics and Astronomy, University of Utah, Salt Lake City, Utah 84102, USA}
\affiliation{Department of Astronomy, University of Science and Technology of China, Hefei, Anhui 230026, China}

\author[0000-0003-0706-8465]{Yi-Fu Cai}
\affiliation{Department of Astronomy, University of Science and Technology of China, Hefei, Anhui 230026, China}
\affiliation{Institute of Deep Space Sciences, Deep Space Exploration Lab, Hefei, Anhui 230026, China}

\author{Di He}
\affiliation{Department of Astronomy, University of Science and Technology of China, Hefei, Anhui 230026, China}

% \author{Ao Wang}
% \affiliation{CAS Key Laboratory of Theoretical Physics, Institute of Theoretical Physics, Chinese Academy of Sciences, Beijing 100190, China}

\author{Qingqing Wang}
\affiliation{Department of Astronomy, University of Science and Technology of China, Hefei, Anhui 230026, China}

\author{Ruoxi Yang}
\affiliation{Department of Astronomy, University of Science and Technology of China, Hefei, Anhui 230026, China}
%\author{et al}
%\affiliation{Astronomy Department, University of Science and Technology of China, \\
%NO.96 Jinzhai Road, Hefei Anhui 230026, China \\
%}

\begin{abstract}
\nolinenumbers
We propose a novel microlensing event search method that differs from either the traditional time domain method, astrometric microlensing, or the parallax microlensing method. Our method assumes that stars with nearly identical “genes" - normalized Spectral Energy Distributions (SED) bear the same luminosity within the intrinsic scatter due to stellar properties. Given a sample of stars with similar normalized SEDs, the outliers in luminosity distribution can be considered microlensing events by excluding other possible variations. In this case, we can select microlensing events from archive data rather than time domain monitoring the sky, which we describe as static microlensing. Following this concept, we collect the data from Gaia DR3 and SDSS DR16 from the northern galactic cap at high galactic latitudes. This area is not preferable for normal microlensing search due to the low stellar density and, therefore, low discovery rate. By applying a similarity search algorithm, we find $5$ microlensing candidates in the Galactic halo.  
\end{abstract}

\section{Introduction} \label{sec:intro}
Gravitational lensing is a general relativity phenomenon in which the light from a background source is bent and focused by the gravitational field of a lens object \citep{Dyson1920RSPTA, Einstein1936Sci}. Micro-gravitational lensing (microlensing, \cite{1986ApJ...304....1P}) refers to gravitation lensing when the lens object is massive and compact. The lens passes between an observer and a distant background source star, creating a gravitational field strong enough to bend the light from the source and magnify its flux as it travels toward the observer \citep{1996ARA&A..34..419P}. It is particularly useful for faint object detection because it does not depend on the light emitted by the lens object itself. Instead, it uses light from the background objects to probe the foreground object, providing information about the mass, velocity, and distance of the lens object combined via light curve time duration $t_{\rm E}$ \citep{1996ApJ...473...57M}. This unique property makes it possible to detect objects that have little or no electromagnetic emission, such as black holes, MACHO (MAssive Compact Halo Object, a dark matter candidate), and exoplanets \citep{2012ARA&A..50..411G}. However, the probability of such events is rare due to the small cross-section between two stellar objects in our Galaxy.

Despite the rarity of the microlensing phenomenon, it has been blossoming with the help of large time-domain surveys. In the  1990s, there were three first-generation microlensing surveys, namely the MACHO project \citep{Alcock2000ApJ}, the EROS (Expérience pour la Recherche d'Objets Sombres) project \citep{Tisserand2009A&A}, and the OGLE (Optical Gravitational Lensing Experiment) project \citep{Udalski1997AcA}. Then followed by many other surveys targeting microlensing events, for instance,  MOA (Microlensing Observations in Astrophysics) \citep{Sumi2003ApJ}, SuperMACHO which is a successor of MACHO project \citep{Becker2005IAUS}, KMT \citep{Jung2019AJ} and a very recent one-Subaru/HSC Andromeda observations \citep{Niikura2019NatAs}.  Follow-up surveys are also designed aiming at discovering the exoplanet systems from the newly identified microlensing events based on existing surveys, e.g. PLANET (Probing Lensing Anomalies NETwork) \citep{Albrow1998ApJ} which joined the RoboNet-1.0 \citep{Burgdorf2007P&SS} in 2005 and merged with the MicroFUN (Microlensing Follow-Up Network) project in 2009, an informal consortium of observers dedicated to photometric monitoring of interesting microlensing events in the Galactic Bulge \footnote{https://cgi.astronomy.osu.edu/microfun/}.

These surveys yield numerous scientific discoveries ranging from the constraints on the upper limit of MACHO or PBH (primordial black hole) fraction as dark matter \citep[e.g.][]{Alcock1998ApJ, Niikura2019PhRvD}, to rouge (free-floating) black holes \citep{Sahu2022ApJ} or intermediate massive black hole \citep{Mirhosseini2018A&A} and to the structure of inner Milky Way \citep{Gyuk1999ApJ}. Noticeably, a recent work by \cite{Lin2022arXiv221100666L} constrained the size of the dark matter core of the Milky Way halo by fully analyzing the two-dimensional microlensing event rate sky map based on the newly released OGLE-IV data \citep{Mroz2019ApJS}, their result supports a dark flat core with the size of about $300$pc. 

% Surveys that are not originally designed for microlensing, however, the time domain data are natural gold mines for such detection. 

Nevertheless, most of the microlensing surveys have two natural limitations. Firstly, the preferred survey regions focus on the galactic bulge, disk, the Large Magellanic Cloud, and M31 where the star densities are high enough to enhance the microlensing detection probability. Secondly, different surveys have various observing cadences, leading to different efficiency \citep{Gaudi2000ApJ}. To design an efficient microlensing survey, a selected dense stellar background and daily or hourly-based cadence are preferred. Moreover, the length of surveys limits the discovery of long-duration lensing events, which are crucial to identifying massive objects such as intermediate black holes.

Although typical microlensing surveys stare at dense regions (e.g. the galactic center) with a high cadence, other time-domain imaging surveys with longer cadence also provide chances to find microlensing events in different places.
Gaia \citep{Gaia2023A&A} which has been charting the three-dimensional map of the Milky Way for a decade, detected 363 microlensing events between 2014 and 2017 \citep{Wyrzykowski2023A&A}. The ZTF (Zwicky Transient Factory) \citep{ZTF2019PASP} with 47 square degrees field of view, discovered 19 out of 60 microlensing events beyond the galactic plane ($|b|\geq 10^o$) and 1558 microlensing candidates have been recorded. 
LSST (Legacy Survey of Space and Time) \citep{Winch2022ApJ} for the southern hemisphere and WFST (Wide Field Survey Telescope) \citep{Wang2023SCPMA} for the northern hemisphere will be the most powerful time-domain survey machines, collecting terabytes of data per night. They will enable the detection of many microlensing events, leading to a promising era expected for microlensing research. 
However, they have longer cadences than typical microlensing surveys, so the challenge of how to utilize the data for microlensing searching emerges. We thus need a method complementary to the traditional light curve method, especially for long cadence surveys. 

There are already methods developed to detect the microlensing event other than using light curves, e.g. astrometric microlensing \citep{Nucita2017IJMPD} as well as lensing parallax \citep{Gawade2024MNRAS}. However, these methods are strongly limited by the requirement of high-accuracy astrometry, allowing only application on time series Gaia data.

In this work, we thus propose a novel microlensing detection method, \textbf{static microlensing}, which enables us to dig from any existing archived photometric data and spectroscopic data. Instead of comparing the brightness of the same object at different epochs, we use the spectrum as a fingerprint of a specific star to create an ensemble of stars that are intrinsically similar to the specific star, then compare the luminosity of the specific star to the luminosity distribution of the ensemble. The basic idea is similar to the spectroscopic parallax method in the cosmic distance ladder, but we use distance to infer luminosity rather than use the luminosity to infer distance in the spectroscopic parallax. In this work, we develop an efficient pipeline to construct the ensemble and detect the microlensing events, then we validate the pipeline with mock data. Finally, we apply the technique to the large sample of the stellar spectra and magnitudes from the Sloan Digital Sky Survey (SDSS) and stellar parallax from Gaia. 

% The basic idea is similar to the "twin paradox" that 

% \begin{itemize}
%     \item Introduce the importance of microlensing detection and weakness
% \end{itemize}
% \section{Method} \label{sec:micro}
% \begin{itemize}
%     \item Introduce the concept of static microlensing
%     \item Introduce the Faiss algorithm and k-means algorithm
%     \item the confirmation method
% \end{itemize}

% In this work, combining the large sample of the stellar spectrum from Sloan Digital Sky Survey (SDSS) and Gaia DR3, we design a new microlensing candidate searching method so-called ``Static Microlensing''. In this method, we compare the absolute luminosity in a sample of similar stars selected by their spectrum and pick out the significantly brighter outlier from other stars as our microlensing candidates. 
The paper is organized as follows. Section \ref{concept_validation} describes the concept and algorithm of static microlensing and mock results. Section \ref{sec:data} introduces stellar catalogs and how stellar samples are constructed for the static microlensing searching pipeline. Section \ref{sec:res} shows our microlensing candidates and discussion of the results. We finally give a conclusion in \ref{sec:conclusion}.

\section{Static Microlensing Concept, algorithm and mock event}\label{concept_validation}

In this section, we detail the concept of static microlensing and potential detection ability via a set of realistic mock data.

\subsection{Static Microlensing Concept}
Traditional microlensing utilizes the light curves, i.e. flux variation, of background stars from time domain observations to capture the time-varying lensing magnification.
% or the change of parallax of the lensed star with stringent conditions \citep{Alcok1995ApJ} as well as the recent astrometric microlensing \citep{Nutita2017IJMPDa}. 
The static microlensing technique we propose here, nevertheless, can select microlensing events from archived star catalogs with high-quality multi-band photometry, high-resolution spectral energy distributions (SEDs), and high-quality imaging without time domain observations.

The theoretical foundation of static microlensing is that stars with similar spectra bear similar intrinsic luminosity.
As confirmed by decades of success of the stellar evolution theory, we can categorize stars into several classes by their colors and absolute magnitudes with the Herzprung-Russel diagram (HR diagram). Stars in each class are similar in physical properties such as temperature and luminosity. The main sequence, the most significant feature on the HR diagram, further provides a strong relation between absolute magnitude and color, enabling distance measurement by photometric parallax. Assuming that stars with similar color bear similar intrinsic luminosity, photometric parallax measures distance modulus $\mu$ by comparing the apparent magnitude $m$ from observation and the absolute magnitude $M$ inferred from colors
\begin{equation}
    \mu^{\rm PhoPa}=m-M({\rm color}).
\end{equation}
However, the intrinsic scatter of the main sequence is still large. 
Spectroscopic parallax \citep{1914ApJ....39..341A,1916PASP...28...61A,1943assw.book.....M}, though, utilizes spectra instead of colors to infer the intrinsic luminosity, 
\begin{equation}
    \mu^{\rm SpePa}=m-M({\rm spectrum}).
\end{equation}

Similar to spectroscopic parallax, static microlensing also infers the intrinsic luminosity of stars from their spectra. Besides, with geometric distance from parallax measurement, we also calculate an absolute magnitude from apparent magnitude. When no lensing event happens, the luminosity inferred in two ways should be the same. Any difference between them is then a probe to microlensing events
\begin{equation}
    \Delta_{\rm mag}=M(\mu,m)-M(\rm{spectrum}).
\end{equation}

\begin{figure}[h]
    \centering
    \includegraphics[width=0.49\textwidth]{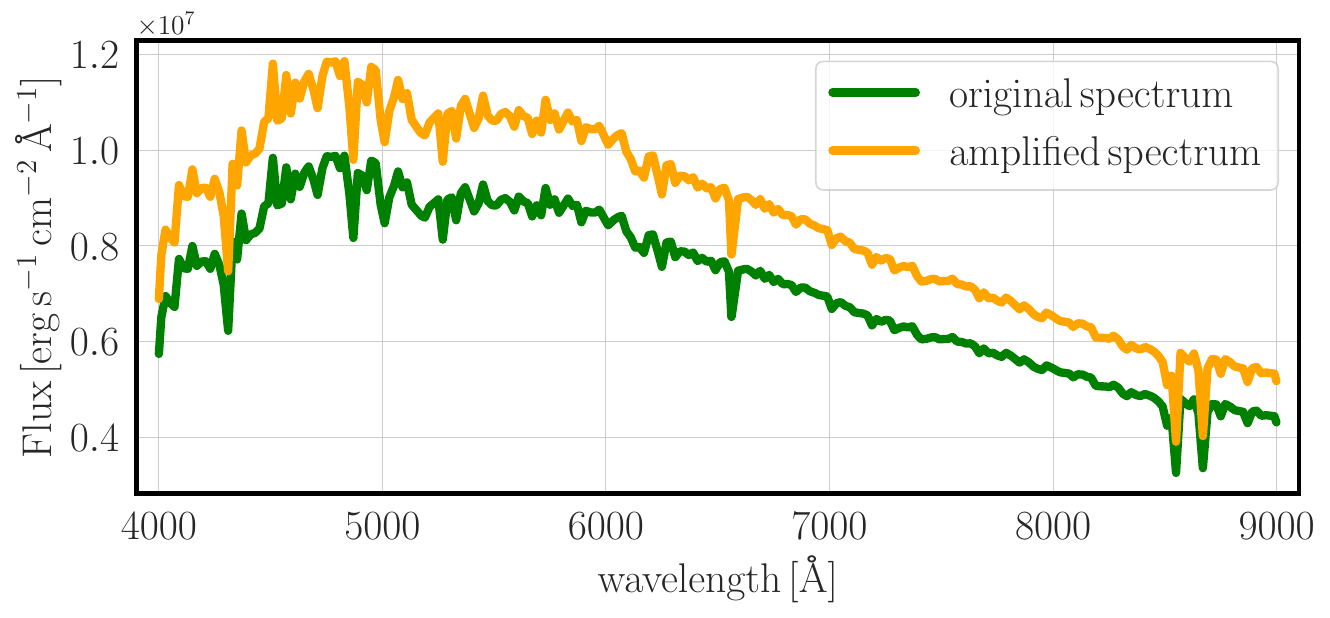}
    \caption{Demonstration of microlensing effect on stellar spectrum between 4000 $\rm \mathring{A}$ and 9000 $\rm \mathring{A}$, with a resolution of 1 $\rm \mathring{A}$. The green line shows the original spectrum while the yellow line shows the amplified spectrum, considering that the microlensing effect is independent of wavelength.}
    \label{fig:concept}
\end{figure}

To avoid strong model dependency, we use an empirical method to infer luminosity from spectra. We construct an ensemble of similar stars by selecting similar spectra. Because of the rarity of microlensing events, we treat the luminosity distribution of the ensemble from parallax and apparent magnitude as the unlensed luminosity distribution for the given spectrum.
The brighter outliers from the luminosity distribution can be considered as microlensing events if we can clearly exclude other intrinsic variability factors, such as flare, etc. The elongation caused by microlensing compared to the nearby unlensed stars can be an extra and effective proof of the microlensing effect and contains more information about the lensing system given a high resolution imaging observation. Fig.~\ref{fig:concept} demonstrates the idea that for two stars with identical SEDs, the yellow SED is magnified by the microlensing effect across the whole spectrum. As a result, the observed lensed star flux is enhanced in all broad-band observations. Here we list four basic requirements for the detection of static microlensing events from the archived data:
\begin{itemize}
    \item a star catalog with high-resolution spectroscopic information;
    \item high accurate multi-band photometry with parallax measurement of those same stars due to the independence of wavelength of lensing magnification;
    \item a reliable selection method to classify those stars into sub-classes with luminosity distribution as narrow as possible;
    \item design a blind tests to check the probability;
    \item (optional) high image resolution to measure the ellipticity of the microlensing candidates and compare them to the nearby stars.
\end{itemize}

In the following sections, we test our method based on mock data focusing on the first three conditions. We will address the fourth issue, high resolution imaging of the elongation of the star image due to the microlensing effect in an independent paper (\textcolor{blue}{He et al in prep}) that focuses on breaking the degeneracy of the lens mass and other parameters in microlensing formulation. This will leads to reliable estimation of the mass of the mass-gaps \cite{Lam2022ApJ} as well as intermediate massive black hole \citep{Mirhosseini2018A&A}.

\subsection{Algorithm}\label{subsec:method&alg}

\begin{figure*}[ht!]
\centering
\includegraphics[width= \textwidth]{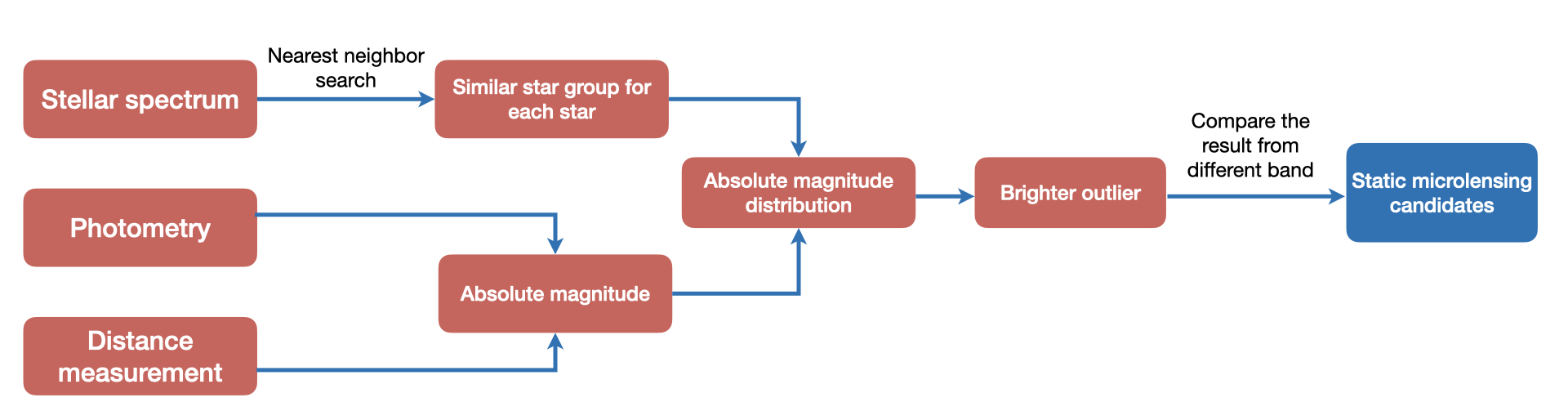}
\caption{A flow chart briefly illustrating our static microlensing searching algorithm.}
\label{fig:method}
\end{figure*}

The key issue in selecting static microlensing event candidates is comparing the absolute magnitude of each target star to other stars with similar physical properties, meaning that we should find neighbors of target stars in the physical property parameter space. To get such star samples, we focused on comparing their stellar spectra which contain information about stars. In this work, we select the nearest (spectrum) vectors in comparison to the target star, which is a way for similarity search.

The idea of similarity search is that for a set of d-dimensional vectors $x_i$ and a specific definition of distance between different vectors, for each vector we can search for its nearest, 2nd nearest,...,$k$-th nearest neighbor based on their distance. In practice, we adopted a nearest neighbor search algorithm library Facebook AI Similarity Search (FAISS)\footnote{\url{https://github.com/facebookresearch/faiss}} which is an open-source library for similarity search and clustering of dense vectors \citep{johnson2019billion}. It provides several vector index construction methods for highly efficient similarity search. In our work, as stellar spectra are simple one-dimensional vectors, we directly calculate their Euclid distance between each vector without any extra encoding. This method is slowest in comparison to other algorithms in the library but gives the highest precision.

Applying FAISS to our stellar spectra samples, for each star, we search for reasonable numbers of nearest spectra to construct a unique stellar sample, which comes to be an absolute magnitude distribution of the target star and its similar neighbors. Then we compare its absolute magnitude with the average value of its neighbor. For all $g$ $r$ $i$ bands, if 
$M_{\rm tar}<\langle M \rangle-3\sigma$ where $ M_{\rm tar}$ is the absolute magnitude of the target star, $ \langle M \rangle$ is the mean absolute magnitude of the cluster and $\sigma$ is the standard deviation of absolute magnitude distribution, the target star is regarded as an initial candidate for static microlensing events. 
% We set an Euclid distance limitation as a upper limit for each star to as far as possible distinct the target from other classifications of stars. A lower limitation of search number is also set to ensure reliable statistical properties. This limitation will be further discussed in the Appendix \ref{app:faiss}.

Considering other intrinsic stellar properties such as variables and flares can also cause optical variability, we further compare the result in different bands, which will be discussed in detail in Section \ref{subsec:optical_var}. 
For each stellar distribution, if 
\begin{equation}
    \Delta_{\mathrm{mag}} = M_{\rm tar}-\langle M \rangle
\end{equation}
are the same within uncertainty 
\begin{equation}
\sigma_f=\sqrt{\frac{\sigma_{\rm tar}^2+\sigma_{\langle M \rangle}^2}{2}},
\end{equation}
where $\sigma_{\rm tar}$ 
contains measurement uncertainty of parallax of target stars and $\sigma_{\langle M \rangle}$ contains measurement uncertainties for the neighbors, we exclude such candidates as a result of flare. Moreover, we exclude candidates with $\sigma_f$ in any band larger than a setting criterion for higher precision. 

For further confirmation, we could combine the images of candidates. Integrating the ellipticity of stellar images caused by microlensing is also a potential way to do further selection. We leave further discuss to future work.

\subsection{Mock events}

\subsubsection{Mock data}\label{mock}

We use the TRDS version of Kurucz 1993 Models\footnote{Kurucz Models:\url{https://www.stsci.edu/hst/instrumentation/reference-data-for-calibration-and-tools/astronomical-catalogs/kurucz-1993-models}} for our mock spectrum. The Kurucz 1993 Atlas contains about 7600 stellar atmosphere models covering a wide range of metallicities ($ \log{Z}$), effective temperatures ($ T_{\rm eff}$), and surface gravity ($ \log g $). The Atlas includes models of metal abundances relative to solar of+1.0, +0.5, +0.3, +0.2, +0.1, +0.0, -0.1, -0.2, -0.3, -0.5, -1.0, -1.5, -2.0, -2.5,-3.0, -3.5, -4.0, -4.5, and -5.0, with models covering the gravity ranging from $ \log{g}$= 0.0 to +5.0 in steps of +0.5. The range of effective temperature is from 3500 K to 50000 K. We also used \texttt{pysynphot}\footnote{pysynphot:\url{https://pysynphot.readthedocs.io/en/latest/}} to directly get the absolute flux for any parameter. Other spectra corresponding to different parameter spaces from above are the interpolation of the existing spectra.

For the mock spectra, We computed the \texttt{gri} absolute magnitudes observed from SDSS referring to \cite{Tokunaga2005}. For simplicity, we don't take into consideration of airmass (i.e. assuming airmass=0). In detail, for each band, we get interpolated SDSS filter response $\mathrm{R}(\lambda)$ first from SDSS filter response functions. Then, we get the pivot wavelength $\lambda_{\mathrm{eff}}$ from
\begin{equation}
    \lambda_{\mathrm{eff}}^2 = \frac{\int \lambda \mathrm{R} \mathrm{d}\lambda}{\int (\mathrm{R}/\lambda)\mathrm{d}\lambda},
\end{equation}
and mean photon rate density from
\begin{equation}
    \langle f_\lambda \rangle  = \frac{\int \lambda \mathrm{R} f_\lambda \mathrm{d}\lambda}{\int \mathrm{R}\lambda\mathrm{d}\lambda},
\end{equation}
where $f_\lambda$ is the flux at wavelength $\lambda$. Finally, we convert the flux to AB magnitude by
\begin{equation}
    M = -2.5\log_{10}\frac{\langle f_\lambda \rangle \lambda_{\mathrm{eff}}^2}{3631[\mathrm{Jy}]c},
\end{equation}
where c is the speed of light and the computed magnitude is nearly SDSS magnitude.

\subsubsection{Mock test}\label{subsubsec:mocktest}
\begin{figure*}[ht!]
    \centering
    \includegraphics[width=\textwidth]{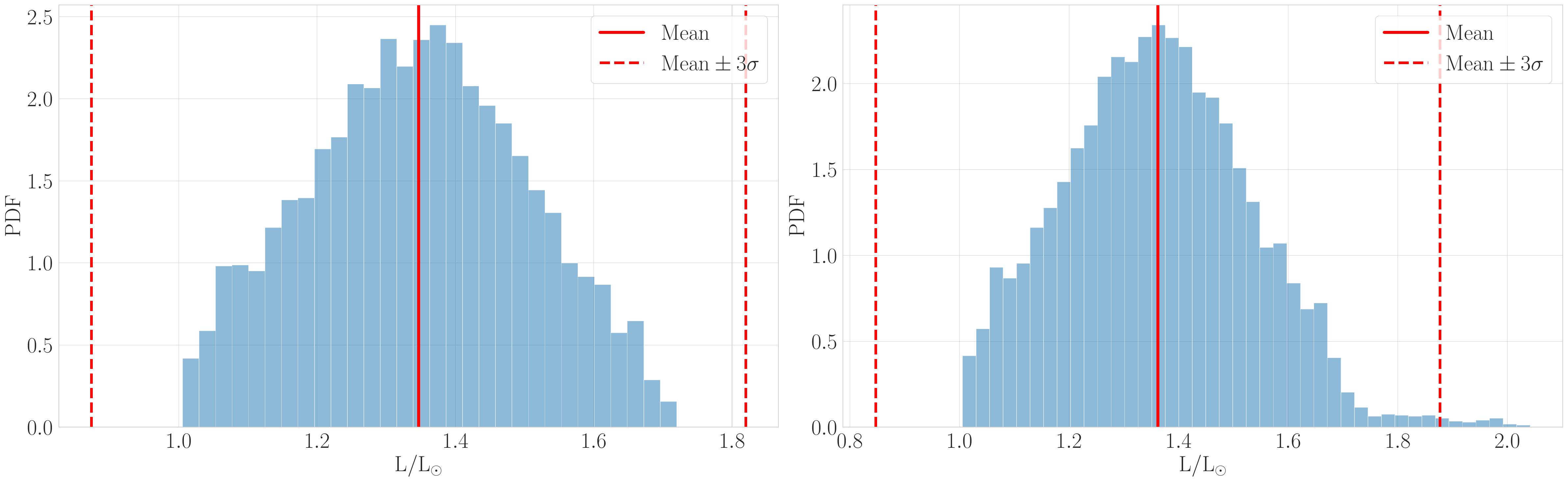}
    \caption{Left panel: Distribution of luminosity of mock samples bearing similar intrinsic properties (i.e. $M_*, Z_{\rm int}, T_{\rm eff}, \log g $). The red solid and dashed lines individually show the mean value and $3\sigma$ interval of the whole distribution. The distribution is nearly Gaussian distribution. Right panel: We randomly choose $\sim 5\%$ of all samples from the distribution and amplify the luminosity by $20\%$. The distribution shows a tail to the brighter side. For some samples, its luminosity goes beyond a $3\sigma$ interval of the distribution, due to the mock microlensing effect here.}
    \label{fig:mesa_lum}
\end{figure*}

First, we use Modules for Experiments in Stellar Astrophysics (MESA)\footnote{MESA:\url{https://docs.mesastar.org/en/release-r23.05.1/}} \citep{Paxton2011}, which is a suite of open source libraries for a wide range of applications in computational stellar astrophysics, including stellar evolution. Here we generate a bunch of main sequence stars, which we defined as $ m_{\rm He}^{\rm core}>0$, where $ m_{\rm He}^{\rm core}$ is the He core mass of a star, given initial conditions including initial mass and initial metallicity ($ Z_{\rm int}$). Given initial mass and $ Z_{\rm int}$, we get the properties including stellar mass, $ T_{\rm eff}$, $ \log g$, and luminosity. For simplicity, we define $ \theta=\{M_*, Z_{\rm int}, T_{\rm eff}, \log g \}$. In our test, the initial mass was set to be uniformly distributed between $[M_\odot,2M_\odot]$ in the step of 0.05 with random noise, while for each mass, the initial $Z$ was set to be uniformly distributed between [0.02,0.04]. We don't use spectra in this part because of the limitation of MESA, which is not able to generate data of spectra. On the one hand, these properties could provide some conditions to get the spectra from the Kurucz model mentioned above. On the other hand, we switch to another side and do a simple test here first. We seek for the similarities of other properties, here $\theta$, and then compare the luminosity for those with near $\theta$. We present a sample (or cluster) in Figure. \ref{fig:mesa_lum}, showing the distribution of the luminosity of these stars. The luminosity distribution is nearly Gaussian without outliers. In the right panel, we randomly choose $\sim 5\%$ of all samples from the distribution and amplify the luminosity by $20\%$, which significantly change the luminosity distribution.
%the solid red line is the mean value of luminosity, and two dashed red lines show the $3\sigma$ deviation from the mean value. 

\begin{figure*}[ht!]
\centering
\includegraphics[width=\textwidth]{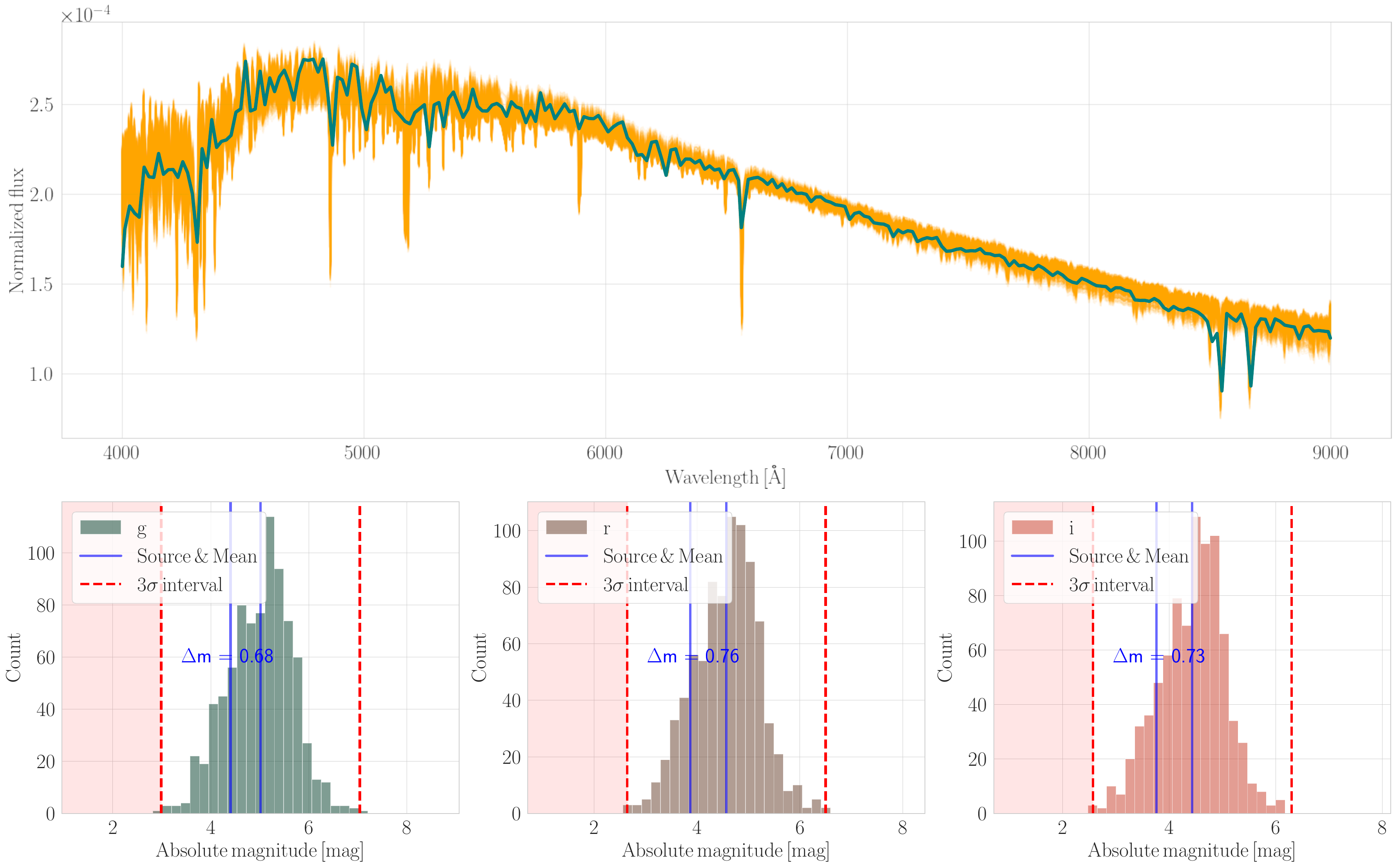}
\caption{Upper panel: The similarity searching results for mock spectrum target. The green line is the mock spectrum and the orange lines represent the stars with similar spectra to the candidate. Similarity search succeeds in finding spectra bearing similar properties to the target. For all neighbor spectra, the black body continuum spectra is quite similar to that of the target. Some spectra show similar absorption lines to the target, which tend to have similar metallicity to the target.\\
Bottom panel: Distribution of absolute magnitude in \textit{gri} bands for the mock sample and its neighbors. We show the mean value and $3\sigma$ region of the distribution in blue and red dashed lines separately. The blue line also shows the magnitude of the mock sample, which is within the $3\sigma$ interval of the distribution.}
\label{fig:kurucz_cla}
\end{figure*}

Then we use spectra from Kurucz model in order to show the basic idea of similarity search. In this part, we integrate spectra from SDSS, combining model spectra and observed spectra together to expand the amount of dataset. 
We select a target spectrum from the Kurucz model with $ Z=0.2$, $ T_{\rm eff}=5750$, and $ \log g=1$ and get absolute flux from \texttt{pysynphot}. 
To make accordance with the spectra of SDSS, we also interpolate to this spectrum between $4000 \rm \mathring{A}$ and $9000 \rm \mathring{A}$ with the steps of 1 $\rm \mathring{A}$ and select similar spectra from SDSS. Here we use \texttt{KDTree} under a random threshold of maximum distance 0.00075. When applying \texttt{KDTree} to search for near spectra, we also tried to set different maximum distances and found that there is not a huge change to the overall result. However, in this part, we choose a larger distance than that using purely observational data considering the noise of observational spectra. Figure. \ref{fig:kurucz_cla} shows the spectra of these stars. The green line shows the normalized spectra of the target star and the orange lines are the normalized spectra of stars selected from SDSS. Based on the distance of the flux array, we selected the spectra with similar spectra, even with some absorption lines. The magnitude distributions of $g r i$ bands are also shown in Figure \ref{fig:kurucz_cla}, in which the blue lines are the mean value of the distribution and the absolute magnitude of the target star. The dashed red lines show a $3\sigma$ deviation from the mean value. The red-shaded region is the potential place where static microlensing events may be located. For $g r i$ bands, the magnitudes of the target star are 4.40, 3.86, and 3.76 respectively after applying the filter of SDSS and integrating over the spectrum. With an amplification of $\delta=0.76$, the magnitude of the target will be beyond the $3\sigma$ region of the overall distribution. It will be regarded as an outlier and thus a candidate for micro-lensing events.

\clearpage
\section{Data}\label{sec:data}

Following the description above, this study requires a sample encompassing large amount of stars, each characterized by high-resolution spectral data and precise distance measurements. Therefore, our sample construction initiates from the cross-matched catalog between Gaia DR3(\citealp{prusti2016gaia},\citealp{vallenari2023gaia}) and Sloan Digital Sky Survey (SDSS) photometric data \citep{marrese2022gaia}. Here we briefly introduce these two datasets and our sample construction process.

\subsection{Gaia DR3}
\label{gaiadr3}
Gaia is a space-based optical telescope developed by the European Space Agency (ESA) to obtain precise astrometric measurements for the objects in our sky\citep{2016}. The Data Release 3 (DR3) of Gaia was published on 13 June 2022 \citep{vallenari2023gaia}. Using the trigonometric parallax method, Gaia DR3 can achieve micro-arcsecond accuracy in distance measurements, improving parallax precision by 30\% over Gaia DR2.  This dataset provides astrometric measurements including positions, parallax, and proper motion for over 1 billion stars with a limiting magnitude of 21 in the G band, as well as astrophysical parameters such as effective temperature ($ T_{\rm eff}$), surface gravity ($ \log g$), and $[M/H]$. Additionally, It provides mean BP/RP spectra for over 2 million objects. However, due to the low resolution of these spectra($R\sim$100 to 30 for BP and 100 to 70 for RP), we instead use the high-resolution spectra from the SDSS dataset, accessed via the cross-matched catalog available in the Gaia dataset \citep{marrese2022gaia}.

\subsection{SDSS DR16}
\label{sdssdr16}
We derived the photometric data and high-resolution spectra from the Sloan Digital Sky Survey (SDSS).
The SDSS is an all-sky spectroscopic and imaging survey, containing optical spectroscopic observations through August 2018 and photometry in \texttt{ugriz} bands from imaging data. SDSS has classified its sources into different class types such as stars and galaxies so we select stars in the cross-matched catalog according to the type label. As for the photometric part, we combine the apparent magnitude from different bands, extinction computed following \cite{Schlegel_1998} with conversion from $\rm E(B-V)$ to total extinction following \cite{2011ApJ...737..103S}, and parallax from Gaia to calculate the absolute magnitude for each star.

For the SDSS spectroscopic data, their spectral resolution $R$ ranges from 1850 to 2200 and covers a range within 3800$ \rm\mathring{A}$ - 9200$\rm\mathring{A}$. We use the modeled spectra to get more accurate results, which have excluded the contribution of atmospheric emission lines. For comparison, we first excluded spectra whose minimum wavelength was higher than $4000 \rm\mathring{A}$ or whose maximum wavelength was less than $9000 \rm\mathring{A}$, then applied interpolation to each spectrum between $4000 \rm\mathring{A}$ and $9000 \rm\mathring{A}$ with the steps of $1 \rm\mathring{A}$ as the coverage of wavelength for each spectrum is different. We finally re-normalized each spectrum by its total flux, which also tends to eliminate the effect caused by micro-lensing for stellar spectra. 

%Also, we excluded spectra whose minimum wavelength was higher than $4000 \rm\mathring{A}$ or whose maximum wavelength was less than $9000 \rm\mathring{A}$ to make the selected sources easier to compare. 

%Before directly classifying these spectra, we make corrections to the spectra considering the extinction and reddening. First, we re-normalized the spectra considering their redshift and apply interpolation to each spectrum between $4000 \rm\mathring{A}$ and $9000 \rm\mathring{A}$ with the steps of $1 \mathring{A}$ as the coverage of wavelength for each spectrum is different. 
% We correct the flux at each wavelength by a factor given by the Milky Way extinction map (\cite{Schlegel_1998} dust map) and CCM89\citep{osti_5125744} law. 
%Then, we re-normalized each spectrum by its total flux, which also tends to eliminate the effect caused by micro-lensing for stellar spectra. 

\subsection{Sample construction} \label{dataselect}

To build a sample including accurate distance measurement, photometric data, and high-resolution spectrum, we combine the data from GAIA and SDSS. We integrate the parallaxes measured from GAIA DR3, photometric data from SDSS DR13, with spectra from SDSS DR16.

We start from the cross-match catalog between Gaia DR3 and SDSS DR13 data\citep{marrese2022gaia}. To get more accurate results, we exclude sources with more than one best-matched stars. We first selected the objects with highly significant parallax measurement, which means the ratio of parallax and error is larger than 10. To avoid the contamination of dust in the galactic disk, we select the object with the absolute value of galactic longitude larger than $15$. As for the spectrum quality selection, there are \texttt{SN\_MEDIAN\_ALL} values representing the median S/N across all good pixels in a stellar spectrum and \texttt{ZWARNING} values to label bad spectra. We use \texttt{SN\_MEDIAN\_ALL}$\leq$10 criterion to collect the high significant spectrum, and \texttt{ZWARNING}$=0$ flag to discard the contamination from observation. Finally, we derive a sample with $94230$ stellar spectra with absolute luminosity measurement.

Moreover, to minimize the influence of (sub-)giant stars, in this work, we tried to cut the (sub-)giant branch, which will reduce the amount of data but give us a more robust result. We exclude stars with $\rm g-i>1.2$ and $\rm g_{abs}<3.94(g-i)+7.88$ and take the rest of the data (38821 sources left) into consideration. We show the area we study in Figure \ref{fig:outlier_HRD} (Region II) and we will discuss it in section \ref{sec:res}.

\section{Result and Candidates} \label{sec:res}

\begin{figure*}[ht!]
\centering
\includegraphics[width= \textwidth]{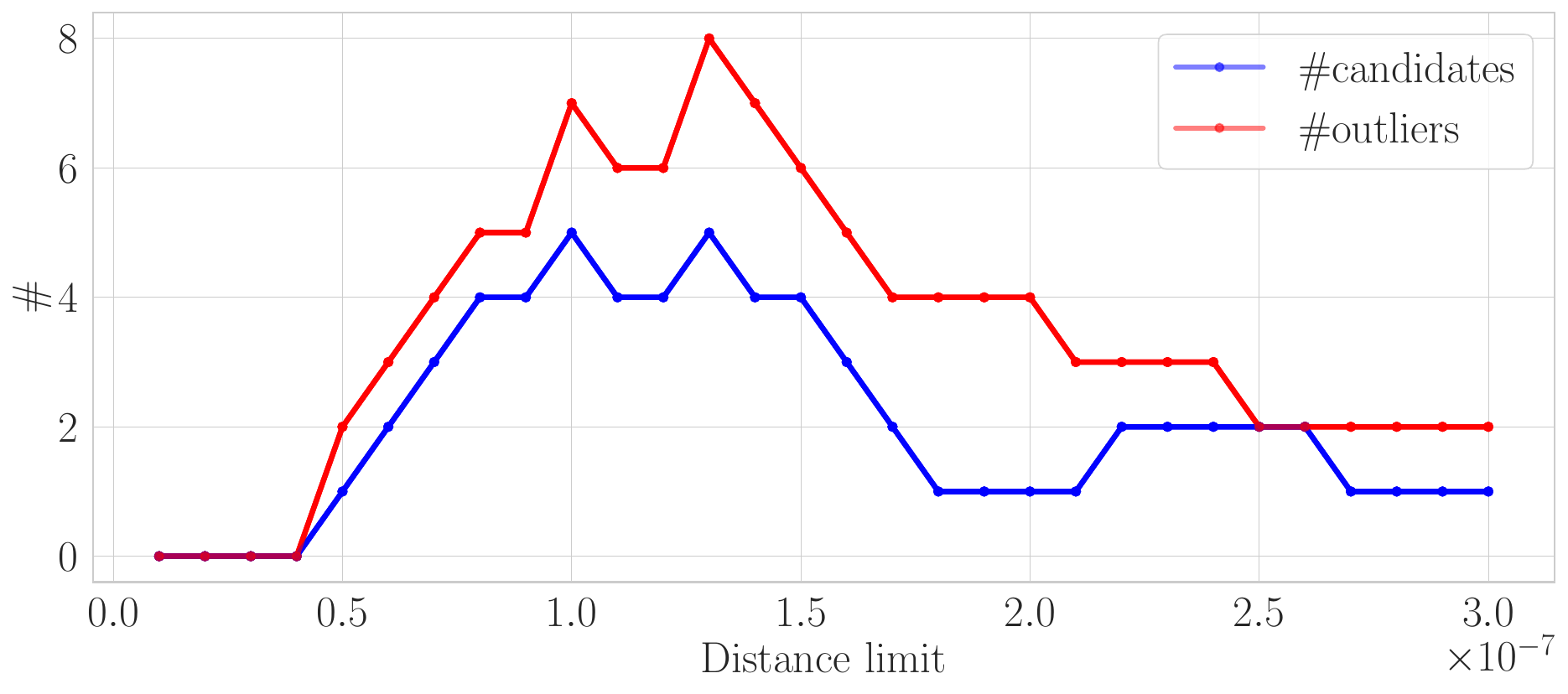}
\caption{The number of outliers (red line) and final candidates after cross \texttt{gri} bands validation (blue line) for different distance limits. To get the distribution of absolute magnitude for statistics, we further set $N_{\mathrm{nei}}>500$. For smaller distance limits, the algorithm cannot select outliers with $N_{\mathrm{nei}}>500$, while for larger distances, the algorithm finds stellar samples whose physical properties are not so similar to the target star. Finally, we choose $1.4\times10^{-7}$ to be the distance limit in this work.}
\label{fig:outlier_num}
\end{figure*}

\subsection{Catalog and Individual Example}
\begin{figure*}[ht!]
\centering
\includegraphics[width=\textwidth]{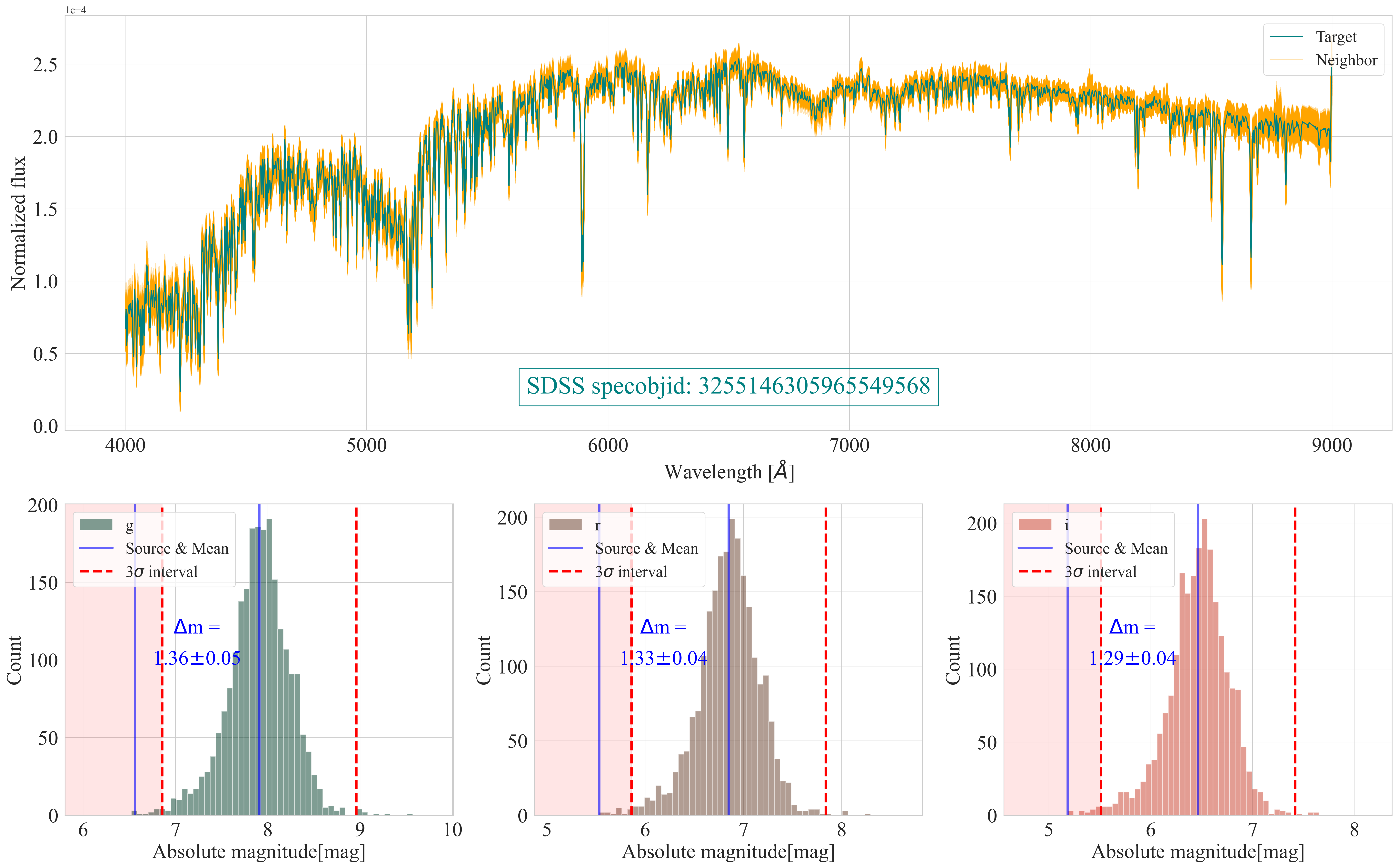}
\caption{Upper panel: An example result of similarity search on our data sample. The green line shows the spectrum of the candidate (specobjid: 3255146305965549568), while the orange lines represent the stars with spectra to the candidate. FAISS do searched for near near stellar spectra for the candidate, some stars even with the same absorption lines.\\
Bottom panel: Distribution of absolute magnitude in \textit{gri} bands. The magnitudes of the candidate are beyond the $3\sigma$ region of the distribution of its family for each band, showing a tendency to be brighter than other stars with similar intrinsic properties.}
\label{fig:spec_example}
\end{figure*}

The original data constructed from the SDSS and Gaia catalog contains various types of stars, including (sub-)giant stars, white dwarfs, and main sequence stars, the latter being our primary focus. Fig.\ref{fig:outlier_HRD} displays the distribution of all original data on a color-absolute magnitude diagram, with sample density indicated by the color map and the positions of our static microlensing candidates marked. Our previous tests showed that if the entire dataset is used as input for our algorithm, the resulting candidates are mostly concentrated on the giant star branch. This result is reasonable, as the possible variability of (sub-)giant stars could introduce contamination into the algorithm. Consequently, we applied the criteria detailed in Section \ref{dataselect} to exclude this branch, with the removed region indicated in Fig. \ref{fig:outlier_HRD}.

We search the nearest stellar spectra for the data using the algorithm described in Section \ref{subsec:method&alg}. The first thing we should be concerned about is the maximum neighbor number limit for each star to search. An absolute number cut is not suitable as the sample density is not uniform, and this may bring bias for different types of stars. Instead, we apply an upper vector distance limitation. A minimum neighbor number limitation should also be set to ensure we have enough neighbors to be compared. To find out the most appropriate vector distance limitation $d_{\mathrm{lim}}$, we try different limitations settings, and the compared result is shown in Fig.\ref{fig:outlier_num}. This figure shows the results when setting $N_{\mathrm{nei}}>500$. After applying comparison of $\Delta_{mag}$ between $g$ $r$ $i$ bands, number of outliers $\# \mathrm{outliers}$ decreases to  number of candidates $\# \mathrm{candidates}$, ruling out possibles intrinsic varibilities like flares. We also tried other $N_{\mathrm{nei}}$ there isn't a huge change in results. In this paper, we use $d_{\mathrm{lim}}=1.4\times 10^{-7}$ and will make the results of other criteria public.

In Figure \ref{fig:spec_example}, which is similar to Figure \ref{fig:kurucz_cla} but the data here are all from observation, we show the result of spectra similarity searching and \texttt{gri} band absolute magnitude distribution for a static microlensing candidate. The upper panel of Figure \ref{fig:spec_example} shows the spectrum of the target star and the spectra of its neighbors. By applying an upper limit for search based on distance, our method successfully derives a sample of the nearest stellar spectra for the target star. The bottom panel shows the distributions of absolute magnitude from all these analogous stars, which are nearly symmetrical unimodal distributions. For these distributions, the magnitude of the target star lies in the left red dashed region, which is beyond the 3$\sigma$ interval for a distribution and means that the target is significantly brighter than its counterparts. We also show the deviation of the target magnitude from the mean value of the distribution, labeled in the figure.

Besides 4 candidates of main sequence (MS) stars, we also select a white dwarf (WD) (SM3 in Table \ref{tab:catalog}). Though the WD candidate isn't a member of the candidates shown in Figure \ref{fig:outlier_num} as it doesn't meet the requirement $N_{\mathrm{nei}}>500$, we still keep it as a candidate due to variabilities of stars. The WD is much fainter (g band magnitude 1.95$\rm ^{mag}$) than other stars (g band magnitude $\sim$ 6.5$\rm ^{mag}$). The number of neighbors of this WD candidate is much lower than other candidates under our vector distance limitation set in the algorithm. Considering this lower number is because of the low WD density in our data sample, we still select this candidate for our final result. This WD candidate also indicates that our algorithm can find out possible micro-lensing events as a white dwarf generally does not have intrinsic luminosity variability.

% \begin{deluxetable*}{lrr}
% \caption{The Header of Complete Catalog}
% \label{tab:catalog}
% \tablehead{
% \colhead{Column Name}&
% \colhead{Description}&
% \colhead{Format}
% }
% \startdata
% gaia\_source\_id&Object ID in Gaia DR3                      & \\
% sdss\_objid     &Object ID in SDSS DR13 photometric catalog &\\
% sdss\_specobjid &Spectrum Object ID in SDSS                 &\\
% ra              &Right Ascension (J2000.0)                  &decimal degrees\\
% dec             &Declination (J2000.0)                      &decimal degrees\\
% l               &galactic longitude                         &decimal degrees\\
% b               &galactic latitude                          &decimal degrees\\
% parallax        &stellar parallax in Gaia DR3               &milliarcsecond\\
% parallax\_error &standard error of stellar parallax         &milliarcsecond\\
% psfmag\_X       &PSF magnitude in band X(g,r,i)             &asinh mag\\
% psfmagerr\_X    &PSF magnitude error in band X              &asinh mag\\
% extinction\_X   &extinction in band X                       &asinh mag\\
% abs\_psfmag\_X  &absolute mag in band X                     &asinh mag\\
% dis\_mean\_X    &mean value of the magnitude distribution   &asinh mag\\
% dis\_mean\_err\_X  &error of the mean value                 &asinh mag\\
% dis\_std\_X     &standard error of magnitude distribution   &asinh mag\\
% delta\_X        &deviation from the mean value              &asinh mag\\
% delta\_err\_X   &error of the deviation                     &asinh mag\\
% note        &  & \\
% \enddata
% \end{deluxetable*}

\begin{table*}
\label{tab:catalog}

\centering

\begin{tabularx}{\textwidth}{rrrrrr} 
% \begin{tabularx}{\textwidth}{llllll} 

\toprule
ID        & SM1 & SM2 & SM3  & SM4 & SM5\\
\midrule
 Gaia Source ID & \makecell[l]{681241633\\451076224} & \makecell{211054810 \\ 1784126336} & \makecell{398664609 \\ 3630282112$^\star$} & \makecell{153323309 \\ 7662303616} & \makecell{117598836 \\ 9500059904} \\
% Gaia Source ID & 681241633
% 451076224 & 211054810
% 1784126336 & 398664609
% 3630282112$^\star$ &153323309
% 7662303616 & 117598836
% 9500059904 \\
 SDSS objID & \makecell[l]{123766034 \\ 3387750589} & \makecell{123766873 \\ 5759090903} & \makecell{123766773 \\ 3962227736} & \makecell{123766222 \\ 5141334204} & \makecell{1237662238 \\ 553735321} \\
 % SDSS objID & 123766034\newline3387750589 & 123766873\newline5759090903 & 123766773\newline3962227736 & 123766222\newline5141334204 & 1237662238\newline553735321 \\
 SDSS specobjID & \makecell[l]{325514630 \\ 5965549568} & \makecell{315029962 \\ 4074405888} & \makecell{279001797\\ 1043461120} & \makecell{364578795 \\ 0249254912} & \makecell{373027937 \\ 3442490368} \\
 %SDSS specobjID & 325514630\newline5965549568 & 315029962\newline4074405888 & 279001797\newline1043461120 & 364578795\newline0249254912 & 373027937\newline3442490368 \\
     RA(J2000) & 119.212 & 280.251 & 161.200 & 185.323 & 216.440 \\
     DEC(J2000) & 23.763 & 40.956 & 19.720 & 40.050 & 9.297 \\
     parallax[mas] & $1.411\pm0.041$ & $0.952\pm0.028$ & $7.863\pm0.104$ & $0.999\pm0.034$ & $1.144\pm0.037$ \\
     $\mathrm{T_{eff}[K]}$ & -- & 4518.428 & -- & 4528.848 & 4501.320 \\
     $\log(\mathrm{g})$ & -- & 4.337 & -- & 4.366 & 4.451 \\
     $A_g$ & 0.226 & 0.197 & 0.140 & 0.059 & 0.079 \\
     $M_g$ & $6.560\pm0.012$ & $6.435\pm0.008$ & $11.949\pm0.022$ & $6.654\pm0.017$ & $6.994\pm0.014$ \\
     $\overline{M}_g$ & 7.905 & 7.468 & 13.176 & 7.612 & 7.974 \\
     $\sigma_g$ & 0.351 & 0.316 & 0.376 & 0.296 & 0.312 \\
     \textbf{$\Delta M_g$} & \textbf{1.355} & \textbf{1.050} & \textbf{1.200} & \textbf{0.973} & \textbf{0.993} \\
     $A_r$ & 0.156 & 0.136 & 0.097 & 0.041 & 0.055 \\
     $M_r$ & $5.531\pm0.014$ & $5.539\pm0.006$ & $11.902\pm0.018$ & $5.681\pm0.014$ & $5.918\pm0.016$ \\
     $\overline{M}_r$ & 6.849 & 6.533 & 13.111 & 6.628 & 6.894 \\
     $\sigma_r$ & 0.330 & 0.311 & 0.356 & 0.283 & 0.292 \\
     \textbf{$\Delta M_r$} & \textbf{1.332} & \textbf{1.012} & \textbf{1.191} & \textbf{0.957} & \textbf{0.987} \\
     $A_i$ & 0.116 & 0.101 & 0.072 & 0.031 & 0.041 \\
     $M_i$ & $5.185\pm 0.011$ & $5.234\pm 0.005$ & $11.908\pm 0.021$ & $5.379\pm 0.019$ & $5.570\pm 0.017$ \\
     $\overline{M}_i$ & 6.468 & 6.203 & 13.136 & 6.290 & 6.510 \\
     $\sigma_i$ & 0.329 & 0.304 & 0.348 & 0.299 & 0.297 \\
     \textbf{$\Delta M_i$} & \textbf{1.295} & \textbf{0.991} & \textbf{1.204} & \textbf{0.918} & \textbf{0.948} \\
\bottomrule
\end{tabularx}
\caption{The properties of our Static Microlensing candidate. The parallax, effective temperature, and surface acceleration of gravity are from Gaia measurement. We also present the absolute magnitude $M_x$, the average of absolute magnitude for the candidate and its neighbors $\overline{M}_x$, the standard deviation of magnitude distribution $\sigma_x$ and the magnitude difference between candidate magnitude and average value $\Delta M_x$.}
\end{table*}

\subsection{Properties of the candidates}

In this section, we analyze the properties of these 5 candidates. We separately show the spatial distributions of our static micro-lensing candidates on the galactic extinction map and color-absolute magnitude diagram.

\begin{figure}[ht!]
    \centering
    \includegraphics[width=\linewidth]{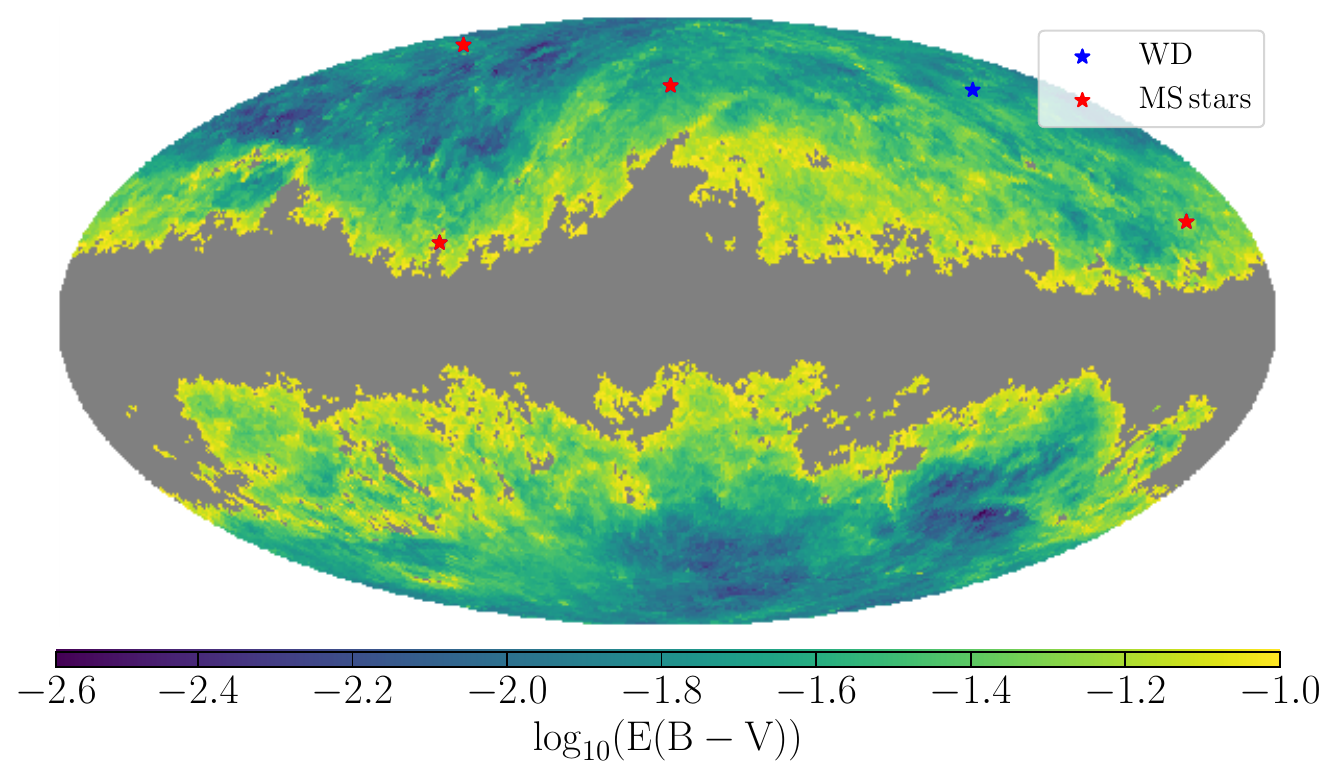}
    \caption{Spatial distribution of candidates on the extinction map where $\rm E(B-V)>0.1$ are masked. In this panel, low latitude regions are masked due to higher dust extinction so we only focus on high latitude stars ($l>15^\circ$). The blue star represents the WD candidate while 4 red stars represent MS star candidates.}
    \label{fig:ext_map}
\end{figure}

As the observational property we focus on is luminosity, we should consider the impact of foreground dust. The corresponding property is the extinction. Figure \ref{fig:ext_map} shows the spatial positions of candidates on the dust extinction map of Milky Way (MW), with $\rm E(B-V)$ from \cite{Schlegel_1998} dust map FITS files. This figure shows regions with $\rm E(B-V)>0.1$ masked and the locations of the WD candidate and MS star candidates. Combined with the extinction values for each band given by the SDSS photometric catalog, the candidates are in regions that are less tend to be influenced by extinction. Moreover, as what we compared with is the deviation of $\rm M_{tar}$ from $\rm M_{mean}-3\sigma$, extinction will not have a large influence on the results. Also, we only focus on stars at high latitudes, so we don't correct stellar spectra due to dust extinction or reddening as high latitude regions host less dust so our results are less tend to be contaminated by extinction.

\begin{figure*}[ht!]
\centering
\includegraphics[width= 0.99\textwidth]{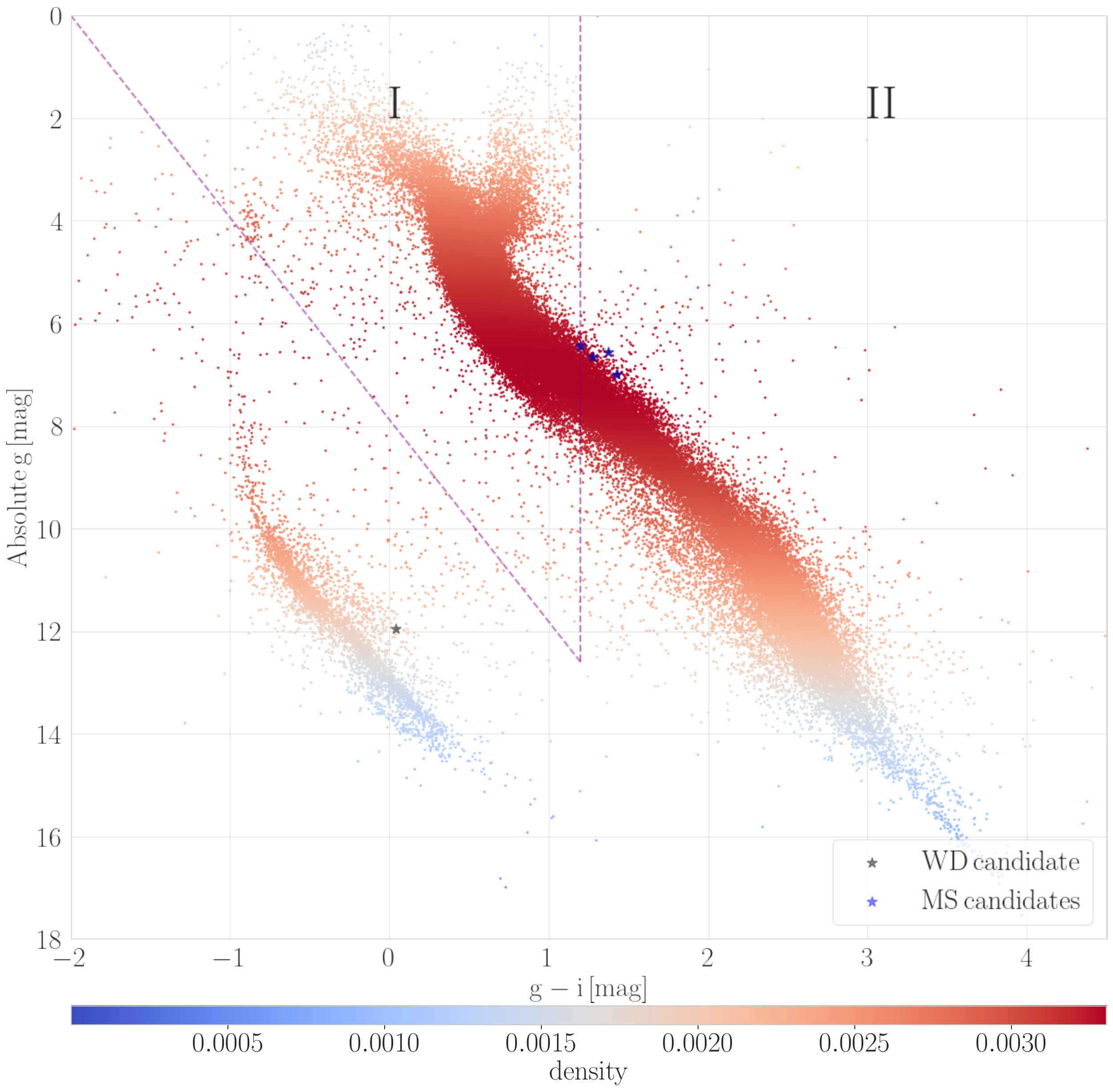}
\caption{Positions of candidates on the CMD. The background shows all data with color representing the density of data points. 
The purple dashed lines show the criteria we apply to exclude the contamination of red giant stars (region I), leaving us with region II data for analysis. We show the candidates after applying the cut with blue point and grey points representing WD candidate and MS star candidates separately. The candidates tend to be located in higher-density regions as there is a higher possibility to find more neighbors for a target star. Compared to other stars with the same $\rm g-i$, the candidates tend to have smaller absolute \texttt{g} band magnitude, showing a tendency to be brighter than other stars with similar intrinsic properties.}
\label{fig:outlier_HRD}
\end{figure*}

We show the distribution of candidates on the color-absolute magnitude diagram in Figure \ref{fig:outlier_HRD}, plotting the WD candidate and MS star candidates separately. We show the original data sample before the giant branch cut in this figure including the main sequence, red giant branch, and white dwarf branch. The background color shows the density of data points, with the majority of data being MS stars. By applying the criteria excluding red giants we only get less than half of the data (from 94230 to 38821), in this way, our results are less tend to be contaminated by variability of (sub-) giants. The candidates tend to appear in high-density regions in the color-absolute magnitude diagram, in which the algorithm can find more neighbors for a target star. Moreover, one candidate is located in the white dwarf branch but is also recognized to be a variable in our algorithm.

\subsection{Validation}\label{subsec:validation}
To make sure the candidates we derive are mostly possible from a microlensing event, there are two problems we should discuss:(i) Does variation of luminosity originate from a microlensing event? (ii) Does the outlier chosen from the nearest spectra search have a similar intrinsic luminosity with its spectra neighbor? 

\subsubsection{optical variability}\label{subsec:optical_var}
Besides microlensing, some other mechanisms would lead to optical variation, such as stellar flares. Flares are stellar energy bursts reported in all spectral classes \citep{kowalski2024stellarflare}. These events can enhance the stellar luminosity several times in a short time and occur from seconds to days \citep{kowalski2024stellarflare}. Flares can be detected in all wavelengths including optical band \citep{pettersen1989flare_review}, so they can contaminate our static microlensing candidates as the photometry catalog we use is derived from just one exposure. However, the stellar spectrum can be changed by flares \citep{kowalski2024stellarflare}, and the variations are varied in different wavelengths. Therefore, we limit the magnitude deviation to the neighbor across \texttt{gri} bands to exclude the possible flare events, to further ensure the variation of the candidates originates from extrinsic mechanisms such as microlensing.

\subsubsection{Candidate variation test}
We compare the absolute magnitude distribution between SDSS and Gaia data for the candidate and its neighboring stars. If the magnitude difference is due to variability rather than inherent physical properties, we would expect the luminosity of candidate to return to the median of the magnitude distribution, given that the two measurements are separated by roughly 10 years.  

In our analysis, we utilized absolute magnitude from the SDSS g-band and Gaia G-band, comparing the position of our candidate within these two distributions. Only the candidate with Gaia source ID $681241633451076224$ exhibited a significant change between these two distributions. The magnitude difference from the candidate magnitude to the average value shifted from $1.36$ to $0.63$.

\subsubsection{efficiency}

\begin{figure}
    \centering
    \includegraphics[width=0.49\textwidth]{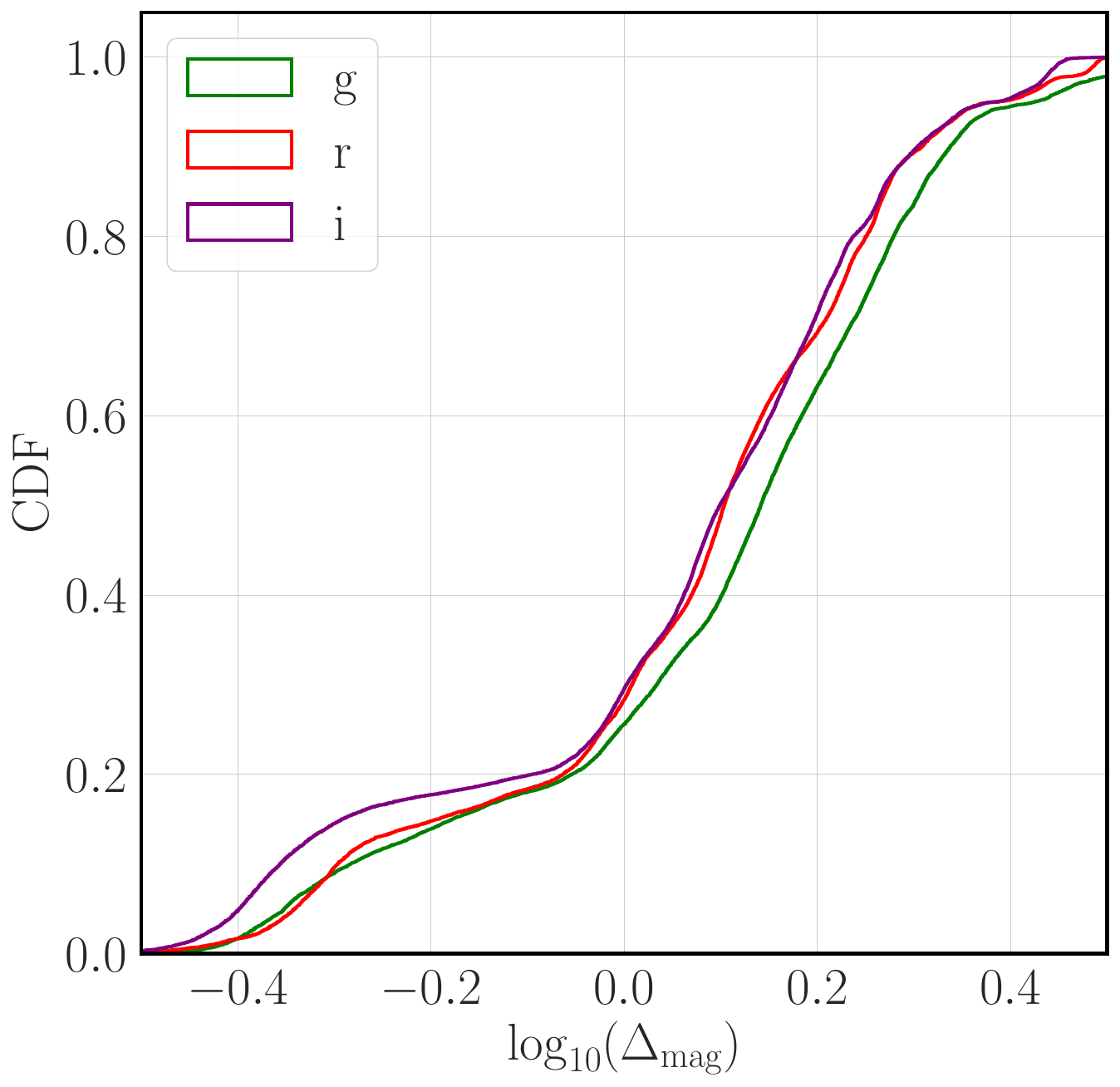}
    \caption{The efficiency curve for \texttt{g}(green) \texttt{r}(red) \texttt{i}(purple) bands. With larger amplification of luminosity, the stars are easier to be selected as static microlensing candidates. When $\log_{10}(\Delta_{\mathrm{mag}})$ is larger than 0.4, the efficiency reaches nearly $100\%$, showing that the algorithm is sensitive to selected such microlensing events.}
    \label{fig:eff}
\end{figure}

For the static microlensing searching algorithm in this work, we estimated the capability of this pipeline and show the efficiency curve in Figure \ref{fig:eff}.

We collected the sources which are not selected out as micro-lensing events (38816 data). For each source, we then computed $\Delta_{\rm{mag}}=M_{\rm{tar}}-(\langle M \rangle-3\sigma)$ in each band and take the value $\Delta_{\rm{mag}}$ as the minimum amplification for this source, i.e. a microlensing event with magnify $\Delta_{\rm{mag}}$ for this source would be detected by the pipeline. We take values of $\Delta_{\rm{mag}}$ as the response (efficiency) of this pipeline to the dataset. We show the cumulative density function (CDF) of $\log_{10}(\Delta_{\rm{mag}})$ in Figure \ref{fig:eff} as the efficiency of this method. We show the efficiency curve for \texttt{gri} bands and there are nearly no differences between these 3 bands. For microlensing events with $\Delta_{mag}>2.51$, the efficiency reaches $100\%$ and the algorithm could nearly find all these events. This method is sensitive to microlensing events caused by large objects such as IMBH, which tends to have large magnifications. However, for the events with a small magnification, the response is low and we tend to rely on time-domain surveys.

\section{Conclusion}\label{sec:conclusion}
In this work, we come up with a new method to search for microlensing events on the basis that for stars with similar physical properties, their luminosity should be close to each other. If a star is remarkably brighter than other similar stars, this phenomenon might be caused by microlensing and we are searching for such events. Different from previous methods, our method is based on high-resolution stellar spectra instead of light curves. 

Firstly we perform tests on mock data to verify the idea of this method. On the one hand, we study the distribution of a family of stars with similar spectra. We use MESA to generate a bunch of data and find that for stars with similar properties (i.e. stellar mass, initial redshift, effective temperature, and surface gravity), the luminosity distribution of these stars is nearly Gaussian. On the other hand, we use spectra generated from the Kurucz 1998 model and search for similar spectra with SDSS data. We find that similarity search works well in finding stars with nearly identical spectra. Moreover, the absolute magnitudes of the model ``stars'' are within $3\sigma$ uncertainties of the magnitude distributions of the constructed ensemble. We further repeat the mock-star tests for different types of stars and find similar results.

From an observational side, for different stars, the microlensing effect will cause different magnifications to their luminosity but will not change their normalized spectra. In such cases, the absolute magnitudes of the target stars might be located beyond $3\sigma$ uncertainties of the magnitude distribution. So we apply the similarity search method to observational data, integrating SDSS and GAIA. For each target star, we used FAISS to search for stars with similar spectra. Then, we select microlensing candidates when all the \texttt{g r} and \texttt{i} band absolute magnitudes are beyond $3\sigma$ uncertainty of the specific magnitude distribution. Also to exclude the contamination caused by flare, we only select out candidates with variance $ \Delta_{\rm mag}$'s are the same for g r and i bands. 

Moreover, microlensing also affects the images of stars, especially the ellipticity. However, the qualities of SDSS images are not good enough for ellipticity measurements so we leave the detailed study of shape measurement and ellipticity analysis to future work. Furthermore, we can have follow-up observations of light curves to further confirm long-duration microlensing events. 

With these results in hand, we can step further in astrophysical research. On the one hand, these candidates provide a potential way to map the matter distribution of the Milky Way, which benefits in understanding the structure of the Milky Way. On the other hand, with more data, we could have more well-rounded static microlensing candidates, which could give us more complete results.

\section*{Acknowledgment}
%\begin{acknowledgements}

   WL acknowledge the support from the NSFC (NO.12192224), the National Key R\&D Program of China (2023YFA1608100, 2021YFC2203100), the 111 Project for "Observational and Theoretical Research on Dark Matter and Dark Energy" (B23042).
  %Gaia

  This work presents results from the European Space Agency (ESA) space mission Gaia. Gaia data are being processed by the Gaia Data Processing and Analysis Consortium (DPAC). Funding for the DPAC is provided by national institutions, in particular the institutions participating in the Gaia MultiLateral Agreement (MLA). The Gaia mission website is https://www.cosmos.esa.int/gaia. The Gaia archive website is https://archives.esac.esa.int/gaia.

  %SDSS
  Funding for the Sloan Digital Sky Survey IV has been provided by the Alfred P. Sloan Foundation, the U.S. Department of Energy Office of Science, and the Participating Institutions. SDSS acknowledges support and resources from the Center for High-Performance Computing at the University of Utah. The SDSS web site is www.sdss4.org.

  SDSS is managed by the Astrophysical Research Consortium for the Participating Institutions of the SDSS Collaboration including the Brazilian Participation Group, the Carnegie Institution for Science, Carnegie Mellon University, Center for Astrophysics | Harvard \& Smithsonian (CfA), the Chilean Participation Group, the French Participation Group, Instituto de Astrofísica de Canarias, The Johns Hopkins University, Kavli Institute for the Physics and Mathematics of the Universe (IPMU) / University of Tokyo, the Korean Participation Group, Lawrence Berkeley National Laboratory, Leibniz Institut für Astrophysik Potsdam (AIP), Max-Planck-Institut für Astronomie (MPIA Heidelberg), Max-Planck-Institut für Astrophysik (MPA Garching), Max-Planck-Institut für Extraterrestrische Physik (MPE), National Astronomical Observatories of China, New Mexico State University, New York University, University of Notre Dame, Observatório Nacional / MCTI, The Ohio State University, Pennsylvania State University, Shanghai Astronomical Observatory, United Kingdom Participation Group, Universidad Nacional Autónoma de México, University of Arizona, University of Colorado Boulder, University of Oxford, University of Portsmouth, University of Utah, University of Virginia, University of Washington, University of Wisconsin, Vanderbilt University, and Yale University.

%\end{acknowledgements}
%\appendix

% \section{Error estimation}\label{app:error}
% To select candidates whose magnifications ($\Delta_{magnitude}$) of g r i bands are the same within errors, we estimate errors for each candidates following propagation of error mentioned below.

% Considering the error of $p$ \textbf{parallax} ($\sigma_p$) and $m$ \textbf{apparent magnitude} ($\sigma_m$), from the definition of absolute magnitude $M$
% \begin{equation}
%     M=m+5+5\log_{10}(\frac{p}{1000\mathrm{pc}})-extinction,
% \end{equation}
% the error of absolute magnitude is then
% \begin{equation}
%     \sigma_{M}^2=\frac{1}{p}\frac{5}{\ln(10)}\sigma_p^2+\sigma_{M}^2.
% \end{equation}

% However, we don't have the magnitude errors of five bands, under which circumstance the error should be $$\sigma_{M}^2=\frac{1}{p}\frac{5}{\ln(10)}\sigma_p^2$$ where p stands for parallax 

% The mean value of the absolute magnitude of a specific class containing n sources is $$\sigma_{mean}=\sqrt{\frac{\sigma_1^2+\sigma_2^2+\cdots+\sigma_n^2}{n}}.$$

% The error between the candidate and mean value is $$\sigma_f=\sqrt{\frac{\sigma_{candi}^2+\sigma_{mean}^2}{2}}.$$

\bibliographystyle{aasjournal}
\bibliography{micro}
\end{document}